# Prediction of overall survival and molecular markers in gliomas via analysis of digital pathology images using deep learning

Saima Rathore[1,2], Muhammad Aksam Iftikhar[3], Zissimos Mourelatos[4]

[1]Center for Biomedical Image Computing and Analytics, Perelman School of Medicine, University of Pennsylvania, PA, USA
[2]Department of Radiology, Perelman School of Medicine, University of Pennsylvania, PA, USA
[3]COMSATS University Islamabad, Lahore Campus, Lahore, Pakistan
[4]Department of Pathology and Laboratory Medicine, Perelman School of Medicine, University of Pennsylvania, PA, USA

**Abstract.** Cancer histology reveals disease progression and associated molecular processes, and contains rich phenotypic information that is predictive of outcome. In this paper, we developed a computational approach based on deep learning to predict the overall survival and molecular subtypes of glioma patients from microscopic images of tissue biopsies, reflecting measures of microvascular proliferation, mitotic activity, nuclear atypia, and the presence of necrosis. Whole-slide images from 663 unique patients [*IDH*: 333 *IDH*-wildtype, 330 *IDH*-mutants, *1p/19q*: 201 *1p/19q* non-codeleted, 129 *1p/19q* codeleted] were obtained from TCGA. Sub-images that were free of artifacts and that contained viable tumor with descriptive histologic characteristics were extracted, which were further used for training and testing a deep neural network. The output layer of the network was configured in two different ways: (i) a final Cox model layer to output a prediction of patient risk, and (ii) a final layer with sigmoid activation function, and stochastic gradient decent based optimization with binary cross-entropy loss. Both survival prediction and molecular subtype classification produced promising results using our model. The c-statistic was estimated to be 0.82 (p-value=4.8x10-5) between the risk scores of the proposed deep learning model and overall survival, while accuracies of 88% (area under the curve [AUC]=0.86) were achieved in the detection of *IDH* mutational status and *1p/19q* codeletion. These findings suggest that the deep learning techniques can be applied to microscopic images for objective, accurate, and integrated prediction of outcome for glioma patients. The proposed marker may contribute to (i) stratification of patients into clinical trials, (ii) patient selection for targeted therapy, and (iii) personalized treatment planning.

1. **Introduction**

Gliomas are major malignant tumors of brain originating from glial cells and exhibit significant molecular, histological, and imaging heterogeneity across and within patients, as well as variable proliferation, which poses several diagnostic and therapeutic challenges [1, 2]. Various types of gliomas cause mutations in genes such as isocitrate dehydrogenase *(IDH)* gene. Additionally, codeletion of chromosomes 1p and 19q (*1p/19q* codeletion) is also seen as a genetic marker for characterization of gliomas. Diffuse gliomas are first classified into one of three molecular subtypes based on *IDH* mutation and *1p/19q* status. The *IDH*-wildtype astrocytoma are characterized by the absence of both *IDH* and *1p/19q* mutation status; *IDH*-mutant astrocytoma are defined by the presence of *IDH*, but absence of *1p/19q*, and oligodendroglioma are defined by the presence of both the mutations, i.e., *IDH* and *1p/19q* [3]. Figure 1 illustrates this taxonomy in a decision tree form, where the leaves represent the three aforementioned types of glioma. Histological grade of gliomas is then determined within each subtype using histologic characteristics ranging from nuclear morphology to higher-level patterns, like necrosis (pseudopalisading or geographic), presence of abnormal microvascular structures, oligodendroglioma component, etc.

Different groups of genetic aberrations, such as *IDH* and *1p/19q*, have varying sensitivity to targeted therapies [4]. Several studies [5-8] have shown that co-deletion of *1p/19q* chromosome arms has prognostic value in estimating positive response of the tumor to chemotherapy and radiotherapy, both, and is correlated with favorable outcome. Similarly, existing literature has revealed favorable prognosis of *IDH*-mutants compared to *IDH*-wildtype patients [9]. Therefore, predicting *IDH* and *1p/19q* status has very important implications for adopting effective treatment regimens for gliomas. An assessment of these markers and expected survival of gliomas at the initial presentation of disease has huge clinical implications.

Determination of *IDH* and *1p/19q* so far has required *ex-vivo* postoperative or biopsy tissue analyses, which undergoes molecular testing. These molecular assays have several limitations: First, they involve destroying the tissue, thereby making the process non-repeatable; second, they invariably capture genomic or proteomic measurements from a small part of the tumor, thereby underestimating tumor heterogeneity; third, they are very expensive and are not available worldwide; finally, tumor resection/biopsy is sometimes limited, such as in cases of inoperable and deep-seated tumors and in post-surgery follow-ups, and consequently, tissue for molecular assessment and analysis is limited in those cases. Measures provided by microscopic assessment of biopsy specimens is therefore be critical for these cases; the microscopic assessment reveals valuable information critical to disease diagnosis and prognosis [10]. Such phenotypic information may encompass collective information represented by the underlying genomic biomarkers. Therefore, analysis of histology images is a widely adopted practice in medical community. However, manual analysis is a highly subjective, tedious and non-repeatable process owing to the tiresome nature of the analysis and high-demanding expertise required.

Computational pathology is a domain of studying microscopic images of a tissue specimen for revealing profound disease characteristics [11]. Computational techniques overcome the limitations posed by manual assessment and also provide a secondary opinion to histologists about disease diagnosis and prognosis. Digitization of tissue slides and the computational techniques developed on those facilitate the: (i) transfer of image-rich pathology data between distant locations for the purposes of research, diagnosis, and education, (ii) solicitation of remote pathology consults without the need to physically ship slides around, (iii) reduce the need for storing glass slides on site and eliminate the risk of slides getting broken or lost, and most importantly (iv) offer a similar assessment of molecular markers for a fraction of the price incurred on genetic testing.

The objective of this study is to develop deep learning approaches to test the hypothesis that overall-survival and clinically-relevant glioblastoma mutations can be predicted using *ex-vivo* digital pathology images. Such approaches, if reliably detect molecular markers and survival using computational methods at the initial presentation of the disease, will help stratification of patients for current and upcoming therapeutic clinical trials. We have employed an advanced deep learning architecture, named deep-residual learning architecture, for analysis of computational pathology images. We hypothesize that quantification of subtle, yet important and spatially complex histology features as extracted from *ex-vivo* digital pathology images is informative and leads to determining molecular tumor characteristics, herein the *IDH* and *1p/19q*, with sufficient sensitivity and specificity on an individual patient basis.

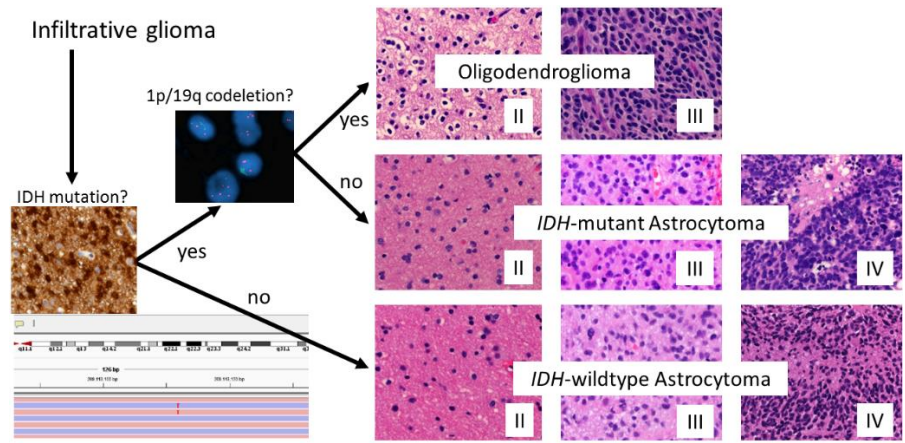

Figure 1: Top-level description of various cancer grades: first classified into one of three molecular subtypes based on *IDH1/IDH2* mutations and the codeletion of chromosomes *1p* and *19q*, *IDH*-wildtype astrocytoma characterized by the absence of both *IDH* and *1p/19q* mutation status, *IDH*-mutant astrocytoma defined by the presence of *IDH* but the absence of *1p/19q*, and oligodendroglioma defined by the presence of both the mutations, i.e., *IDH* and *1p/19q*

## 2. Material and Methods

### 2.1 Distribution of the study cohort

The study cohort includes 663 formalin-fixed, paraffin-embedded glioma specimens of digital pathology images obtained from The Cancer Genome Atlas (TCGA) [12] program. The images were clinically diagnosed with different sub-types of glioma including *IDH*-wildtype, *1p/19q* non-codeleted and *1p/19q* codeleted (Table 2).

Table 1: Class distribution of molecular sub-types of glioma

| Characteristic | IDH WT, n=333 | | | IDH mutant, n=330 | | | | | |
| --- | --- | --- | --- | --- | --- | --- | --- | --- | --- |
| | | | | 1p/19q-codel, n=129 | | | 1p/19q-non-codel, n= 201 | | |
| | II | III | IV | II | III | IV | II | III | IV |
| No. of patients | 14 | 57 | 262 | 69 | 60 | -- | 96 | 88 | 17 |
| OS in days (mean) | 595.64 | 621.21 | 504.05 | 1143.53 | 857.1 | --- | 1158.99 | 975.26 | 1129.58 |
| Sex (Male/Female) | 8/6 | 28/29 | 160/102 | 39/30 | 36/24 | --- | 51/45 | 52/36 | 12/5 |
| Age in years (mean) | 42.14 | 53.52 | 59.12 | 43.97 | 48.21 | --- | 37.33 | 38.77 | 36.76 |

## 2.2 Convolutional Neural Networks

In the recent era, increasingly advanced and sophisticated computational techniques are being developed for automated medical image analysis. This highly attributes to the continuous development of imaging technologies and wider availability of high-power computing machines. Deep learning techniques represent the range of methods, which have recently attracted researchers' interest owing to the availability of massive amounts of data along with the above-mentioned factors. Models based on deep learning paradigm have surpassed their competitive models from classical machine learning [13], which focus on finding predictive patterns from engineered features [9, 14, 15]. This advantage is attributed to the inherent ability of deep learning models to learn the features capable of discerning the patterns of interest. Due to their strong predictive ability, many deep learning models have been successfully applied to medical applications, such as cancer detection [16-18], localization [19], and grading [20].

An artificial neural network (ANN) is a biologically-inspired computational modelling tool, which comprises layers of interconnected compute nodes. These computing nodes, called artificial neurons, loosely simulate that of a biological neuron and are arranged in different layers. Each artificial neuron in a layer receives some input signal for processing and passes the output signal to other connected nodes in the next layer. Such networks of artificial neurons learn unknown patterns from example data using the back-propagation principle. For example, different sub-types of glioma can be identified using ANN by inputting a multitude of example cases from these sub-types. Generally, an ANN may consist of 3 layers namely input, hidden and output layer. However, in many applications, the layers in the network may be arbitrarily large in number. Such a network is termed as a 'deep' neural network or simply a deep learning model.

Convolutional Neural Network (CNNs) is a special type of deep neural network applicable to image data, which applies convolution operations to input image in different layers of the network. A number of convolutional filters are applied at each layer for transforming the input image into predictive feature-maps. These feature-maps are produced as a result of a training process, which adjusts convolutional filter parameters using the back-propagation principle on the prediction error. The feature-maps encode significant image characteristics predictive of specific patterns, and are robust to noise in the image. Many recent medical applications have employed CNN for producing remarkable accuracy [21, 22].

## 2.3 The ResNet architecture for glioma prediction

Deep learning models are designed on the philosophy of making the model as 'deep' as the validation loss can be decreased. However, the depth of network contributes to the vanishing gradient problem, thereby overfitting to the training data. ResNet deep learning architecture has been proposed in this context to solve the problem of vanishing gradients by introducing skipped connections. This paper investigates the application of ResNet model for accurate discrimination of different sub-types of glioma and survival prediction using digital pathology images of tissue biopsies.

## 2.4 Data curation and training setup of the ResNet architecture

Digital pathology images were obtained for each patient, and 100 regions of interest (ROI) were delineated for each tissue image. The ROIs were selected in a way to make sure that they contained via tumor area, descriptive of characteristics of tumor.

The ResNet architecture (Figure 2) is used to detect the sub-type of glioma as well as patient survival. The ROIs extracted from histologic images were given as input to the ResNet architecture. The convolutional layers of the ResNet architecture produced feature maps, which were indicative of the descriptive characteristics of different sub-types of glioma, and the max pooling layers resampled the intermediate feature maps to reduce the complexity of the feature space. Followed by convolutional operators, vector space embedding was performed, that converted the feature maps into vectorized format. The vectorized maps were then fed to fully connected layers, where several non-linear transformations were applied on the feature maps to learn the characteristics of gliomas and to adjust system weights accordingly. These fully connected layers were finally connected to the output layer, which was configured in two different ways: First, the final layer was setup with sigmoid activation function and binary cross-entropy loss to produce a binary output. The final prediction label of a sample was computed by majority voting of the 100 constituent ROIs for the categorical problem where positive and negative labels were generated by the sigmoid function. Second, the output layer was modeled as a Cox-layer, which produced the patient risk of death. The final prediction label of a sample was computed by calculating the median value of the risks of 100 constituent ROIs. Experimental results in terms of correlation coefficients, c-statistics, area under the curve, and accuracy for survival outcome prediction and molecular subtypes, respectively, showed the effectiveness of deep learning techniques for the purpose.

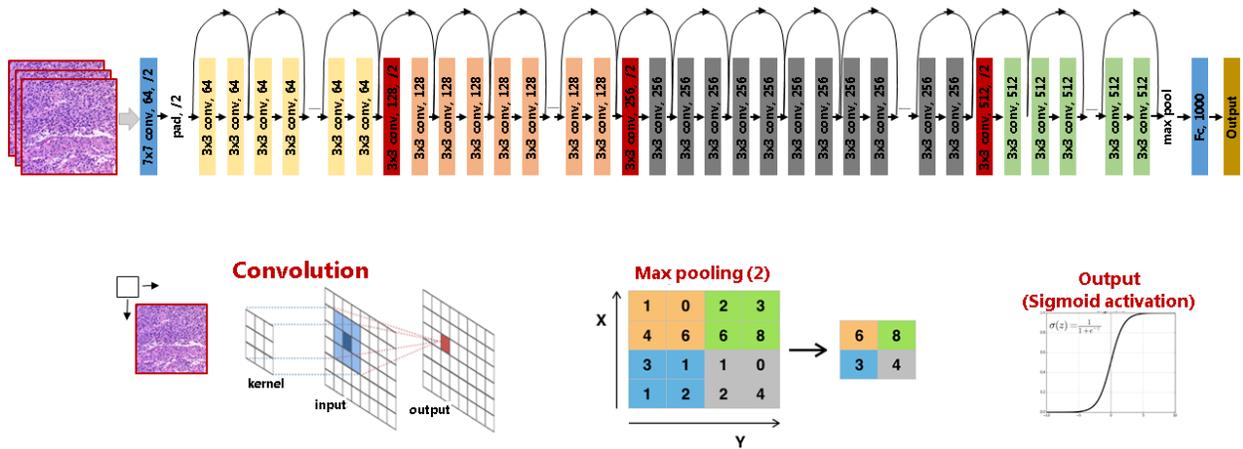

Figure 2: The ResNet architecture for glioma sub-types prediction, including mutant vs wildtype IDH, codeleted vs non-codeleted 1p/19q, and short vs long survivors

### 3. Experimental Results

Performance of our classification schemes in predicting wildtype vs. mutant *IDH*, codeleted vs non-codeleted *1p/19q*, and long vs short overall-survival are presented in Table 2. The quantitative results were obtained for all grades to validate the generalizability and robustness of the trained model in different disease severities. The ROC curves for each of our predictive schemes in stratification of patients based on their *IDH*, *1p/19q* or survival are also shown in Figure 3.

### 3.1 Classification Performance of Predictive Model in Detecting IDH Mutation Status

Our predictive model's accuracy in correctly classifying the *IDH* status on complete dataset was 88.92% (sensitivity = 87.77%, specificity = 84.35%). The model's accuracy in correctly predicting *IDH* status within different histologic groups was 90.50% (sensitivity = 91.52%, specificity = 78.57%), 91.21% (sensitivity = 91.89%, specificity = 89.47%), and 92.77% (sensitivity = 77.78%, specificity = 93.51%), respectively. Assessment of the mutation status was estimated in four different experiments, each time using 25% testing dataset. The models were trained and validated using 50% and 25% datasets, respectively, whereas the performance was evaluated in the left out 25% cohort after independently applying the model trained in the training cohort. Furthermore, a receiver operating characteristic (ROC) analysis was performed and the area under the curve (AUC) for each ROC was 0.86, 0.90, 0.90, and 0.90 for complete dataset, grade-II, grade-III and grade-IV, respectively.

Table 2: Quantitative results for IDH (+/-) classification experiments. The experiments were performed in the complete dataset as well as in each cancer grade, i.e. Grade-II, Grade-III and Grade-IV.

| | Accuracy | Sensitivity | Specificity | AUC |
|---|---|---|---|---|
| IDH mutant/wildtype | | | | |
| All | 88.92 | 87.77 | 84.35 | 0.86, 0.01 [0.84-0.89] |
| Grade II | 90.50 | 91.52 | 78.57 | 0.90, 0.02 [0.84-0.95] |
| Grade III | 91.21 | 91.89 | 89.47 | 0.90, 0.02 [0.86-0.94] |
| Grade IV | 92.77 | 77.78 | 93.51 | 0.90, 0.05 [0.80-0.99] |
| 1p/19q codeleted/non-codeleted | | | | |
| All | 88.23 | 87.38 | 88.58 | 0.86, 0.02 [0.82-0.91] |
| Grade II | 91.51 | 91.30 | 91.66 | 0.89, 0.02 [0.83-0.94] |
| Grade III | 92.56 | 93.33 | 92.04 | 0.91, 0.02 [0.86-0.97] |
| Survival short/long (Classification model) | | | | |
| All | 84.32 | 83.23 | 84.76 | 0.86, 0.06 [0.75-0.92] |
| Grade II | 88.43 | 87.34 | 89.13 | 0.90, 0.03 [0.81-0.96] |
| Grade III | 90.89 | 92.67 | 84.12 | 0.87, 0.04 [0.80-0.93] |
| Grade IV | 91.32 | 91.26 | 90.08 | 0.91, 0.01 [0.83-0.97] |

### 3.2 Classification Performance of Predictive Model in Detecting 1p/19q Mutation Status

The next experiment discriminates between *IDH* mutant astrocytoma (*1p/19q* co-deleted) and oligodendroglioma (*1p/19q* non-co-deleted). Our predictive model's accuracy in correctly classifying the *1p/19q* status on complete dataset was 88.23% (sensitivity=87.38%, specificity=88.58%). The model's accuracy in correctly predicting *1p/19q* status within different histologic groups was 91.51% (sensitivity=91.30%, sensitivity=91.66%), and 92.56% (sensitivity=93.33%, specificity=92.04%), respectively. Assessment of the mutation status was estimated in four different experiments, each time using 25% testing dataset. The models were trained and validated using 50% and 25% datasets, respectively, whereas the performance was evaluated in the left out 25% cohort after independently applying the model trained in the training cohort. Furthermore, AUC for each ROC was 0.86, 0.89, and 0.91 for complete dataset, grade-II, and grade-III, respectively. The results are not shown for grade IV, as the TCGA dataset does not include grade IV IDH mutant astrocytoma cases, so the classification between IDH mutant astrocytoma and oligodendroglioma was not possible.

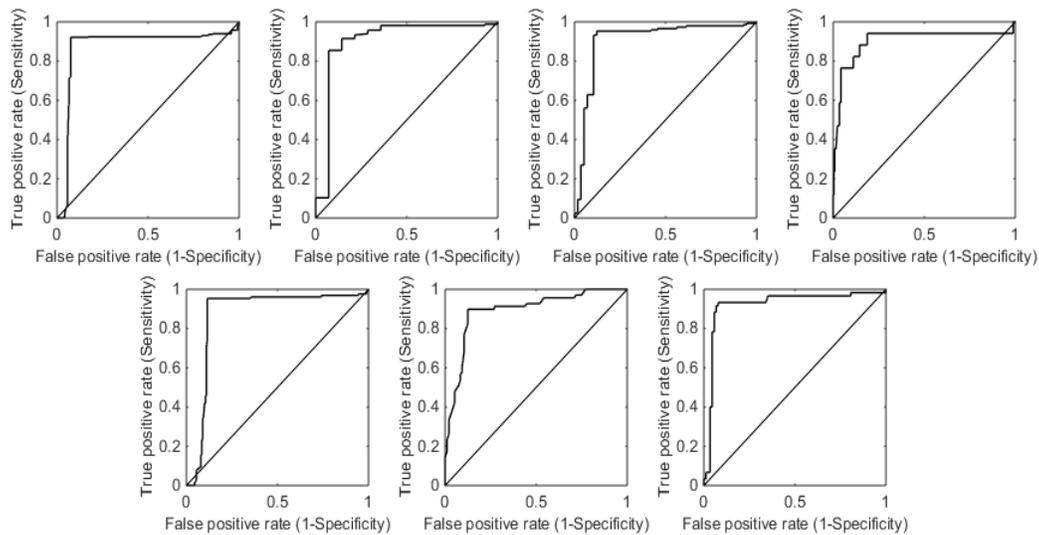

Figure 3: Top row: ROC curves for IDH (+/-) classification experiments: left to right show ROC curves for complete dataset, Grade-II, Grade-III and Grade-IV, respectively. Bottom row: ROC curves for 1p/19q (codeleted/non-codeleted) classification experiments: left to right show ROC curves for complete dataset, Grade-II, and Grade-III, respectively. ROC curves were not evaluated for 1p/19q detection in grade IV group owing to the very small number of patients in that group.

### 3.3 Classification Performance of Predictive Model in Detecting Long and Short Survivors

The regression models developed within different molecular subtypes, i.e., *IDH* wild-type astrocytoma, *IDH* mutant astrocytoma, and oligodendroglioma, respectively, were able to yield correlation scores of $\rho$=0.81 (AUC=0.79), $\rho$=0.80 (AUC=0.78) and $\rho$=0.80 (AUC=0.81), respectively (Figure 4). The concordance index (c-index), respectively, was 0.85 ($p < 0.01$), 0.81 ($p <0.01$) and 0.79 ($p < 0.01$) within *IDH* wild-type astrocytoma, *IDH* mutant astrocytoma, and oligodendroglioma, respectively. The risks predicted by the model, shown in Figure 4 bottom-row) correlate with both histologic grade (Grade-II, III, and IV) and molecular subtype (IDH wild-type astrocytoma, IDH mutant astrocytoma, and oligodendroglioma), decreasing with grade and generally trending with the clinical aggressiveness of genomic subtypes. A classification model was also developed for the prediction of survival, and corresponding results are given in Table 2.

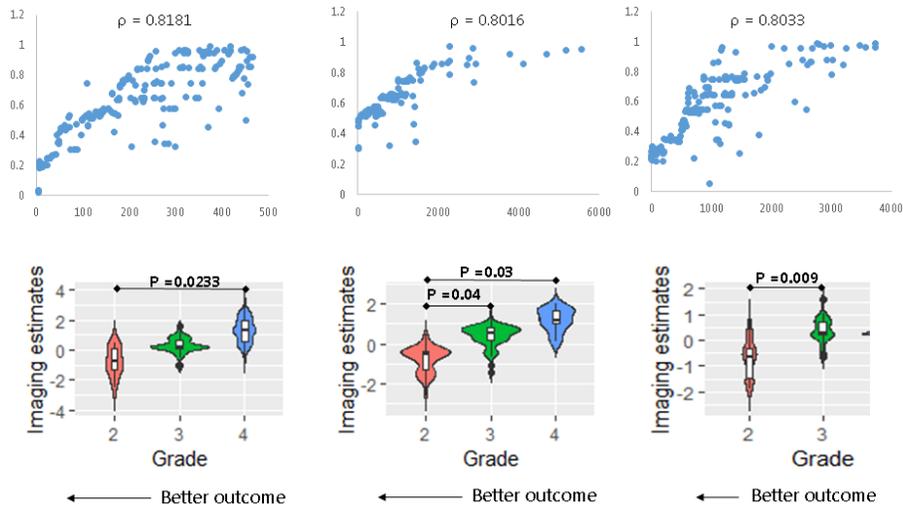

Figure 4: Top row: Scatter plot showing the predicted and the true values of overall-survival (in days). Bottom Row: The risks predicted by the models correlate with both histologic grade and molecular subtype, decreasing with grade and generally trending with the clinical aggressiveness of genomic subtypes. Left-to-right: Different molecular subtypes, i.e., IDH wild-type astrocytoma, IDH mutant astrocytoma, and oligodendroglioma.

## 4. Discussion

In this paper, we have demonstrated accurate prediction of survival outcome and molecular markers for glioblastoma patients using an automated computational pathology pipeline. The method proposed herein involved extraction of 300 patches per image in order to build deep learning models, both for the assessment of survival and molecular markers. Our method consistently provided accurate predictions across all the experiments.

Deep learning architecture employed here aims to skip the feature engineering and make inference based on the raw inputs. The intermediate data-driven features arise from highly non-linear operations. In many medical imaging benchmarks, deep learning methods outperform traditional machine learning methods [23, 24]. These deep learning techniques have facilitated the assessment of multiple imaging and clinical features simultaneously, and are particularly important when analyzing large datasets, as they can identify complex patterns in data that simple regression models obscure. This type of comprehensive approach is critical, as it can learn the patterns which other machine learning algorithms cannot. We believe that our study is the most comprehensive to date; we used over 600 whole-slide images to construct our predictive models.

### 4.1 Validation of the proposed method across multi-institutional dataset

To confirm that our methods would be applicable across multiple institutions, we trained our model on TCIA dataset that has been acquired from multiple institutions under diverse acquisition settings. Our findings (Table 2) support the notion that the deep learning method proposed here in this study allows robust classification of brain tumor datasets arising from multiple institutions, even if a new dataset comes from an institution that was not part of the training sample. The validation of our methods across datasets strongly suggests that our model will perform well in routine clinical settings where samples are much more diverse than in controlled experimental settings.

### 4.2 Importance of the study

The computational pathology pipeline described in this paper addresses many of the current barriers in clinical histopathology. This method offers a standardized approach to resolve intra- and inter-observer variability amongst pathologists and is easily implemented even in low resource settings with available digital scanning tools. Importantly, although this study is focused on brain tumors, the approach presented here is data-driven and could be adapted to other types of cancer with minimal effort. For example, automated analysis may improve grading in neuroendocrine tumors, where treatment decisions are heavily reliant on accurate grading [25]. Similarly, the speed and reproducibility of this method may improve reliability of grading in heterogeneous samples. One additional application may be to identify tumor origin by facilitating comparisons of multiple biopsy specimens from the same patient to determine the morphological concordance between a metastatic site and a primary tumor.

The proposed method underscores the clinical workflow of pathologists by developing predictors on pathology images. In addition to substantially aiding the pathologists in their reporting workflow and clinical decision making, the computational tools could facilitate the development of imaging based supporting diagnostic assays that could allow for improved risk characterization of a disease. The digital slides and the corresponding results provided via computational methods facilitate the real-time transmission of information-rich digital pathology images and results between different facilities for research, diagnostics, and tutoring purposes. This is also suitable for acquiring second-opinion on difficult cases and the option to provide remote consultation without physically shipping tissue slides across different facilities. The digitization of tissue slides may also advance clinical workflow by minimizing the requirement of storing glass slides in bio-banks of pathology departments and decreasing the risk of glass tissue slides getting damaged.

### 4.3 Clinical applicability

This study suggests that deep learning can accurately predict overall survival and molecular markers in data acquired across different institutions. The use of clinically available imaging protocols renders our study likely for immediate translation into the clinic where it can significantly assist in diagnosis, management and treatment of glioblastoma patients. The proposed approach differs from prior literature on the extensiveness of experiments performed, leading to a comprehensive radiomic signature, providing assessments of molecular markers, and clinical outcome of interest.

### 4.4 Limitations and future work

Our study has several limitations. First, the proposed method relies on the extraction of patches from the whole-slide-images, and has not been evaluated directly on the whole-slide images by themselves. However, the method could be tailored, with appropriate modifications, to be used for the whole-slide images. Second, since our method relies on the manual quality inspection of the patches extracted by the software, it is burdensome and time-consuming. Automated patch extraction methods are needed to overcome this barrier. Finally, we have used retrospective data; a prospective dataset comparing our methods to standard histopathological review would lend further validity to our model.

Beyond the scope of this work, there are several important future research directions. Previous studies have used radiology [26] and pathology [27] data separately for survival assessment of brain tumors, and have only used radiology data for assessment of molecular markers [26]. We aim to examine whether the analysis of hand-crafted histologic features, instead of the deep learning features used here, can similarly predict clinical outcomes and other genomic aberrations. A systematic analysis of these characteristics as part of automatic image analysis framework could lead to a better understanding of the relevant cancer biology as well as bring precision and accuracy into assessment and prediction of the outcome of the cancer.